\documentclass[envcountsame,conference,letterpaper]{IEEEtran}

\usepackage{bbm}

\usepackage[tbtags]{amsmath}
\usepackage{amssymb, amsthm}
\usepackage{url}

\usepackage{breakurl}
\PassOptionsToPackage{hyphens}{url}\usepackage[breaklinks]{hyperref}
\usepackage{lineno}
\usepackage{tikz}
\usetikzlibrary{shapes,calc,arrows,automata,decorations.pathmorphing}
\usepackage{xifthen}
\usepackage{xspace}
\usepackage{cleveref}
\usepackage{stackengine}
\usepackage[utf8]{inputenc}
\usepackage{enumitem}
\usepackage{mathtools}
\usepackage[linesnumbered,noend]{algorithm2e}
\usepackage{stmaryrd}
\usepackage{etoolbox}
\usepackage{enumitem}
\usepackage{multirow}
\usepackage{graphicx}
\usepackage{thmtools}
\usepackage{listings}
\usepackage{bm}
\usepackage{array}
\usepackage{tabularx}
\usepackage[location=appendix,appendixsectionname=Proofs]{moveproofs}
\usepackage{wrapfig}
\usepackage{xspace}

\usepackage{microtype}
\everypar{\looseness=-1}
\allowdisplaybreaks[4]
\makeatletter
\renewcommand{\subsection}{\@startsection{subsection}{3}{\parindent}{0ex plus 0.1ex minus 0.1ex}%
  {0ex}{\normalfont\normalsize\itshape\bfseries}}%
\makeatother

\newcolumntype{Y}{>{\centering\arraybackslash}X}

\makeatletter
\let\xx@thm\@thm
\AtBeginDocument{\let\@thm\xx@thm}
\makeatother

\usepackage[
    hyperref=true,%
    backend=bibtex,%
    firstinits=true,%
    mincrossrefs=1000,%
    maxbibnames=99,%
    sortcites,%
    giveninits=true%
    ]{biblatex}

\hypersetup{
    pdftitle={Proving Non-Termination via Loop Acceleration},
    colorlinks=true,
    linkcolor=blue,
    citecolor=olive,
    filecolor=magenta,
    urlcolor=cyan
}

\let\oldland\land
\renewcommand{\land}{\mathrel{\oldland}}
\renewcommand{\epsilon}{\varepsilon}
\let\oldphi\phi
\let\oldvarphi\varphi
\renewcommand{\phi}{\oldvarphi}
\renewcommand{\varphi}{\oldphi}
\renewcommand{\emptyset}{\varnothing}

\renewcommand{\infty}{\omega}

\newcommand{\charfun}[1]{\left\llbracket#1\right\rrbracket}
\newcommand{\constr}[1]{\ \left[#1\right]}
\newcommand{\vect}[1]{\bm{#1}}
\newcommand{\ZZ}{\mathbb{Z}}

\newcommand{\NN}{\mathbb{N}}
\newcommand{\VV}{\mathcal{V}}

\newcommand{\PP}{\mathcal{T}}
\newcommand{\SSS}{\mathcal{S}}
\newcommand{\Params}{\mathcal{P}}

\renewcommand{\AA}{\mathcal{A}}

\newcommand{\fs}[1]{\mathsf{#1}}
\newcommand{\ff}{\fs{f}}
\newcommand{\fg}{\fs{g}}
\newcommand{\fh}{\fs{h}}

\newcommand{\start}{\fs{start}}

\newcommand{\var}[1]{\mathit{#1}}
\newcommand{\tv}{\var{k}}
\newcommand{\true}{\fs{true}}

\newcommand{\lhs}[1]{\lhsop_{#1}}
\newcommand{\head}[1]{\headop_{#1}}
\newcommand{\update}[1]{\updateop_{#1}}
\newcommand{\rhs}[1]{\rhsop_{#1}}
\newcommand{\guard}[1]{\guardop_{#1}}
\newcommand{\cost}[1]{\costop_{#1}}
\newcommand{\target}[1]{\targetop_{#1}}
\newcommand{\temp}[1]{\tempop_{#1}}
\newcommand{\tool}[1]{\textsf{#1}}
\newcommand{\accel}{\textbf{Accelerate}\xspace}
\newcommand{\nonterm}{\textbf{Nonterm}\xspace}
\newcommand{\chain}{\textbf{Chain}\xspace}
\newcommand{\fixedpoint}{\textbf{Fixpoint}\xspace}
\newcommand{\strengthen}{\textbf{Strengthen}\xspace}

\newcommand{\tox}[4][-2pt]{
  \xxrightarrow{}^{\hspace{-1pt}#3}_{\hspace{-1pt}#4}
}

\stackMath
\newcommand\xxrightarrow[2][]{\mathrel{\smash{%
    \setbox2=\hbox{\stackon{\scriptstyle#1}{#2}}%
    \stackunder[0pt]{%
      \xrightarrow{\makebox[\dimexpr\wd2\relax]{$#2$}}%
    }{%
      \scriptstyle#1\,%
    }%
  }}}
\theoremstyle{theorem}
\newtheorem{theorem}{Theorem}
\newtheorem{example}[theorem]{Example}
\newtheorem{definition}[theorem]{Definition}

\SetKw{Break}{break}
\SetKw{Pif}{if}
\SetKwRepeat{Do}{do}{while}
\SetKwIF{If}{ElsIf}{Else}{if}{:}{elif}{else :}{endif}
\SetKwFor{PWhile}{while}{:}{}
\SetKwFor{PFor}{for}{:}{}

\makeatletter
\newcommand{\algorithmstyle}[1]{\renewcommand{\algocf@style}{#1}}
\makeatother
\DontPrintSemicolon

\DeclareMathOperator{\deduceInvariant}{\mathbf{deduceInvariants}}

\DeclareMathOperator{\solve}{\mathbf{solve}}

\DeclareMathOperator{\lhsop}{lhs}
\DeclareMathOperator{\headop}{src}
\DeclareMathOperator{\updateop}{up}
\DeclareMathOperator{\rhsop}{rhs}
\DeclareMathOperator{\guardop}{guard}
\DeclareMathOperator{\targetop}{dest}
\DeclareMathOperator{\costop}{cost}

\DeclareMathOperator{\dom}{dom}
\DeclareMathOperator{\rng}{rng}
\DeclareMathOperator{\proc}{proc}
\DeclareMathOperator{\dec}{dec}
\DeclareMathOperator{\tempop}{\tau}

\Crefname{definition}{Def.}{Def.}
\Crefname{appendix}{Appendix}{Appendix}
\Crefname{example}{Ex.}{Ex.}
\Crefname{theorem}{Thm.}{Thm.}
\Crefname{lemma}{Lemma}{Lemmas}
\Crefname{section}{Sect.}{Sect.}
\Crefname{algorithm}{Procedure}{Procs}
\Crefname{algocf}{Alg.}{Algorithms}
\Crefname{corollary}{Cor.}{Cor.}

\title{Proving Non-Termination via Loop Acceleration\thanks{funded by DFG grant
    389792660 as part of \href{https://perspicuous-computing.science}{TRR~248} and by DFG
    grant GI 274/6}}
\author{\IEEEauthorblockN{Florian Frohn}
   \IEEEauthorblockA{Max Planck Institute for Informatics, Saarbr\"ucken, Germany\\
florian.frohn@mpi-inf.mpg.de}
\and
\IEEEauthorblockN{Jürgen Giesl}
\IEEEauthorblockA{LuFG Informatik 2, RWTH Aachen University, Germany\\
giesl@informatik.rwth-aachen.de}
}

\IEEEoverridecommandlockouts
\begin{document}

\bstctlcite{IEEEexample:BSTcontrol}

% reduce space before and after displaystyle-equations
\setlength{\abovedisplayskip}{5pt}
\setlength{\belowdisplayskip}{5pt}
\setlength{\abovedisplayshortskip}{2pt}
\setlength{\belowdisplayshortskip}{4pt}

\maketitle

\begin{abstract}
  We present the first approach to prove non-termi\-nation of integer
  programs that is based on
  \emph{loop acceleration}. If our technique cannot show non-termination of a loop,
  it tries to accelerate it instead
in order to find paths to other non-terminating loops automatically. The prerequisites for
our novel loop acceleration technique generalize a simple yet effective
non-termination criterion.
Thus, we can use the same program transformations
to facilitate \emph{both} non-termination proving \emph{and}
loop acceleration. In particular, we present a novel invariant inference technique
that is tailored to our approach.
An extensive evaluation of our fully
automated tool \tool{LoAT} shows that it is competitive with the state of the art.
\end{abstract}

\section{Introduction}
\label{sec:introduction}

Proving non-termination of integer programs is an important research topic
(e.g.,
\cite{anant,hiptnt,t2-nonterm,geometricNonterm,amram18,larraz14,rupak08,velroyen,jbc-nonterm,seahorn-term}).
In another line of research,
under-approximating \emph{loop acceleration} is used to analyze
safety
\cite{underapprox15} and runtime complexity \cite{ijcar16}.
Here, the idea is to replace a loop by
code that mimics $k$ loop iterations, where $k$ is chosen non-deterministically.

Many non-termination techniques first search for a diverging configuration and then prove its reachability.
For the latter, loop acceleration
would be useful, as it allows reasoning about paths with
loops without fixing the number of
unrollings. Still, up to now acceleration has not been used for non-termination proving.

To fill this gap, we design a novel loop acceleration technique whose prerequisites
generalize
a well-known non-termination
criterion. This correspondence is of great value: It allows us to develop an under-approximating program
simplification framework that
progresses incrementally towards
the detection of non-terminating loops \emph{and} the acceleration of other loops.

After introducing preliminaries in \Cref{sec:preliminaries}, we present our
approach in \Cref{sec:simplify,sec:inv}.
It eliminates loops via acceleration and chaining,
or by proving their non-termination and
replacing them by a transition to a special symbol
$\omega$. If a loop cannot be eliminated, then we strengthen its guard by synthesizing suitable invariants.
Our approach also handles nested loops by eliminating inner loops before removing outer loops.
Eventually, this leads to a loop-free program where a trace to $\omega$ yields a witness of
non-termination.  So our main contributions are:

\begin{enumerate}[label=(\alph*), align=left, leftmargin=5pt]
\item \label{it:inc} The applicability of existing under-approximating
  loop acceleration techniques is restricted: The
  technique from \cite{underapprox15} is often inapplicable if the loop condition contains
  invariants and the technique from \cite{ijcar16} requires \emph{metering
    functions} which are often challenging to synthesize.
  Thus, in \Cref{sec:simplify}
  we present a novel loop acceleration technique that generalizes \cite{underapprox15} and does not require metering
  functions,
  and we integrate it into
  a program simplification framework inspired by \cite{ijcar16}.
\item \label{it:inv} We combine our approach with a novel
  invariant inference
  technique in \Cref{sec:inv}.
So if the
    prerequisites of our non\-ter\-mination criterion and our acceleration
    technique are viola\-ted,
  then we try to deduce invariants to make them applicable.
\end{enumerate}

\noindent
From a practical point of view, we contribute

\begin{enumerate}[label=(\alph*),resume, align=left, leftmargin=5pt]
\item an implementation in our open-source tool \tool{LoAT} and
\item an
  extensive evaluation of our implementation, cf.\
  \Cref{sec:experiments}.
\end{enumerate}

Finally, \Cref{sec:conclusion} discusses related work and concludes. All proofs can be
found in the appendix.

\section{Preliminaries}
\label{sec:preliminaries}

We denote vectors $\vect{x}$ by bold letters and the $i^{th}$ element of $\vect{x}$ by $x_i$.  \emph{Transitions}
$\alpha$ have the form $\ff(\vect{x}) \tox{c}{}{} \fg(\vect{t}) \constr{\eta}$. The \emph{left-hand side}
$\lhs{\alpha} = \ff(\vect{x})$ consists of $\alpha$'s \emph{source function symbol} $\head{\alpha} = \ff \in \Sigma$ and a vector of
pairwise different variables $\vect{x} \subset \VV$ ranging over $\ZZ$, where
$\VV$ is countably infinite. The set of
function symbols $\Sigma$ is finite and we assume that all
function symbols have the same arity (otherwise one can add unused arguments).
  We use $\VV(\cdot)$ to denote all
variables occurring in the argument.
$\AA$ denotes the set of all \emph{arithmetic expressions} over $\VV$,
i.e., expressions built from variables, numbers, and arithmetic operations like ``${+}$'', ``${\cdot}$'', etc. The
\emph{guard} $\guard{\alpha} = \eta$ is a \emph{constraint}, i.e., a
finite conjunction\footnote{Note that negations can be expressed by negating
  inequations directly, and disjunctions in programs can be expressed using several transitions.} of inequations over
$\AA$, which we omit if it is empty. The \emph{right-hand side} $\rhs{\alpha} = \fg(\vect{t})$ consists of $\alpha$'s
\emph{destination} $\target{\alpha} = \fg \in \Sigma$ and a vector $\vect{t} \subset \AA$. The \emph{substitution} $\update{\alpha} =
\{\vect{x} \mapsto \vect{t}\}$ is $\alpha$'s \emph{update}.

A substitution is a function $\sigma: \VV \to \AA$.  The \emph{domain} of $\sigma$ is
$\dom(\sigma) = \{ x \in \VV \mid \sigma(x) \neq x \}$ and its \emph{range} is defined as $\rng(\sigma) = \{ \sigma(x) \mid x \in
\dom(\sigma) \}$. We sometimes denote sub\-sti\-tu\-tions by sets of key-value pairs $\{y_1
\mapsto t_1, \ldots, y_k \mapsto t_k \}$ or just $\{\vect{y} \mapsto
\vect{t}\}$. Then
each $x \in \VV \setminus \vect{y}$ is mapped to itself.
For every entity $e$, $\sigma(e)$ results from replacing all free variables in $e$ according to
$\sigma$. If $\rng(\sigma) \subset \ZZ$, then $\sigma$ is a \emph{valuation}.
A first-order formula $\phi$  is \emph{valid} if it is equivalent to $\true$. Moreover,
a valuation $\sigma$ is a
\emph{model} of  $\phi$ (or \emph{satisfies} $\phi$,
denoted $\sigma \models \phi$) if
$\sigma$'s domain contains all free variables of $\phi$ and $\sigma(\phi)$ is valid.

An \emph{integer program} $\PP$ is a finite set of transitions.
Their guards restrict the  control flow,
i.e., $\ff(\vect{x}) \tox{c}{}{}
\fg(\vect{t}) \constr{\eta}$ is only applicable if the current valuation of the variables
satisfies $\eta$.

\begin{example}[Integer Program]
  \label{ex:leading}
  Consider the function \lstinline{start}{\normalfont :}
\begin{lstlisting}[language=Python,basicstyle=\small\tt]
    def start(x, y):
        while x >= 0: x = x - y; y = y + 1
        while  y > 0: y = y - x
\end{lstlisting}
  It corresponds to the following integer program:
  \[
    \begin{array}{l@{\qquad\quad}rlll}
      \alpha_1:&\start(x,y) &\to& \ff(x,y) \\
      \alpha_2:&\ff(x,y) &\to& \ff(x-y,y+1) & \constr{x \geq 0} \\
      \alpha_3:&\ff(x,y) &\to& \fg(x,y) & \constr{x < 0}\\
      \alpha_4:&\fg(x,y) &\to& \fg(x,y-x) & \constr{y > 0}
    \end{array}
  \]
  The function symbols $\ff$ and $\fg$ represent the first and the second loop, respectively. The program does not
  terminate if, e.g., \lstinline{x} and \lstinline{y} are initially \lstinline{0}{\normalfont :} After applying the first loop twice, \lstinline{y} is
  $2$ and \lstinline{x} is $-1$, so that the second loop diverges.
\end{example}

\begin{definition}[Integer Transition Relation]
  \label{def:its-relation}
  A term $\ff(\vect{n})$ where $\vect{n} \subset \ZZ$ is a
  \emph{configuration}.  An integer program $\PP$ induces a relation ${\to_\PP}$ on configurations: We have
  $s \tox{k}{}{\PP} t$ if there is
  an $\alpha \in \PP$ and a model $\sigma$ of $\guard{\alpha}$ such that $\VV(\alpha) \subseteq \dom(\sigma)$, $\sigma(\lhs{\alpha}) = s$, and $\sigma(\rhs{\alpha}) =
  t$.\footnote{Throughout the paper, we use ``${=}$'' for semantic (not syntactic)
    equality w.r.t.\
    arithmetic, e.g., ``$\ff(1+2) = \ff(3)$'' holds.} Then we say that $s$ \emph{evaluates} to $t$.
   As usual,  ${\tox{k}{*}{\PP}}$ is the transitive-reflexive closure of ${\tox{}{}{\PP}}$.

  If there is an infinite ${\tox{}{}{\PP}}$-evaluation that starts with $\start(\vect{n})$ where $\start \in \Sigma$ is the
  \emph{canonical start symbol}, then $\PP$ is \emph{non-terminating} and $\start(\vect{n})$ \emph{witnesses
    non-termination} of $\PP$.
\end{definition}

W.l.o.g.,
$\start$ does not occur on right-hand sides.
Otherwise, one can rename $\start$ to $\start'$ and add a transition $\start(\vect{x}) \tox{0}{}{} \start'(\vect{x})$.
A program $\PP$ is
\emph{simplified} if $\head{\alpha} = \start$ for all  $\alpha\in\PP$. So any run of a simplified program has at
most length one.

By definition, integer programs may contain transitions like $\ff(x) \tox{}{}{} \ff(\frac{x}{2})$.
 While evaluations that would not yield integers get stuck (as,
 e.g., $\ff(\frac{1}{2})$ is not a configuration),
 our technique assumes that the arguments of functions are always integers.
Hence, we restrict ourselves to \emph{well-formed} integer programs.

\begin{definition}[Well-Formedness]
  An integer program $\PP$ is \emph{well formed} if for all transitions $\alpha \in \PP$
  and all models $\sigma$ of $\guard{\alpha}$ with $\VV(\alpha) \subseteq \dom(\sigma)$,
  $\sigma(\rhs{\alpha})$ is a configuration.
\end{definition}

To ensure that the program is initially well formed, we just allow integers, addition, subtraction, and
multiplication in the original program.\footnote{One could also allow expressions like
  $\frac{1}{2} \cdot x^2 + \frac{1}{2} \cdot x$ in the
  initial program, as long as every arithmetic expression maps integers to integers.}
While our approach uses program transformations that may introduce
further operations like division and exponentials, these transformations preserve well-formedness.
We formalize our contributions in terms of \emph{processors}.

\begin{definition}[Processor]
  \label{def:proc}
 Let $\infty \in \Sigma$ be a dedicated fresh function symbol.  A \emph{processor} $\proc$ is a partial
 function which maps integer programs to integer programs. It is \emph{sound} if
 the following holds for all $\PP$
  where $\proc$ is defined:
  \[
    \begin{array}{lll}
      \text{if }& \start(\vect{n}) \tox{\omega}{*}{\proc(\PP)} \infty & \text{or }\\
          &\multicolumn{2}{l}{\start(\vect{n}) \text{ witnesses non-termination of }
        \proc(\PP),}\\
      \text{then }& \start(\vect{n}) \tox{\omega}{*}{\PP} \infty & \text{or }\\
                &\multicolumn{2}{l}{\start(\vect{n}) \text{ witnesses non-termination of } \PP.}
    \end{array}
        \]

    \noindent
  If $\proc$ preserves well-formedness, then $\proc$ is called \emph{safe}.
\end{definition}

So we use the symbol
$\infty$ to represent non-termination (and we omit its arguments for
readability):
If we can transform a program $\PP$
into a simplified program
$\PP'$ via safe and sound processors and $\sigma \models \guard{\alpha}$ for some $\alpha \in \PP'$ with $\rhs{\alpha} = \infty$, then
$\sigma(\lhs{\alpha})$ witnesses non-termination of
$\PP$ due to \Cref{def:proc}.

\section{Simplifying Integer Programs}
\label{sec:simplify}

We now present our Contribution \ref{it:inc} by defining suitable
processors.
In \Cref{subsec:invariants}, we introduce the notions of inva\-riants which are the foundation of our
loop acceleration technique, cf.\ \Cref{subsec:acceleration}.
The remaining processors of our approach are used to combine transitions
(\Cref{subsec:chaining}) and to
finally deduce non-termination (\Cref{subsec:non-termination}).

\medskip

\subsection{Invariants}
\label{subsec:invariants}

Our novel loop acceleration technique relies on the following notions of invariants.
Here, ``$\forall \VV(\PP).\ \psi$'' abbreviates ``$\forall (\VV(\psi) \cap \VV(\PP)).\ \psi$'', i.e., the
quantifier binds all free variables of $\psi$ that occur in $\PP$.

\begin{definition}[Invariants]
  \label{def:invariants}
  Let $\alpha \in \PP$.
  If
  \begin{equation}
    \tag{ci}
    \label{eq:consecution}
       \forall \VV(\PP).\
    \guard{\alpha} \land \phi_{ci} \implies \update{\alpha}(\phi_{ci})
  \end{equation}
  is valid, then $\phi_{ci}$ is a \emph{\underline{c}onditional \underline{i}nvariant} of
  $\alpha$. If
  \begin{equation}
    \tag{si}
    \label{eq:consecution-simple}
      \forall \VV(\PP).\
    \phi_{si} \implies \update{\alpha}(\phi_{si})
  \end{equation}
  is valid, then $\phi_{si}$ is a \emph{\underline{s}imple (conditional) \underline{i}nvariant} of $\alpha$.
  If $\phi_{si}$ is a simple invariant of $\alpha$ and
  \begin{equation}
    \tag{md}
    \label{eq:inverse-consecution}
    \forall \VV(\PP).\
     \phi_{si} \land \update{\alpha}(\phi_{md}) \implies \phi_{md}
  \end{equation}
  is valid, then $\phi_{md}$ is \emph{\underline{m}onotonically \underline{d}ecreasing} for $\phi_{si}$ and $\alpha$.
\end{definition}

Recall that $\phi$ is a (standard) \emph{invariant} of a transition $\alpha$ if $\phi$ holds whenever $\alpha$ is applied in a
program run. If such a standard
invariant $\phi$ satisfies
\eqref{eq:consecution}, then $\phi$ is usually called \emph{inductive}.
In contrast to inductive invariants, a conditional invariant $\phi_{ci}$ does not
have to
hold when the control flow reaches $\alpha$, but if it does, then $\phi_{ci}$ still
holds after applying $\alpha$.
Conditional invariants (resp.\ similar notions) are also used in, e.g.,
\cite{velroyen,larraz13,larraz14,borralleras17,brockschmidt15}.
Monotonic decreasingness is converse to invariance: $\phi_{md}$ is preserved
when the effect of $\update{\alpha}$ is undone.

We call constraints of the form $\phi_{ci} \land \phi_{si} \land \phi_{md}$
\emph{monotonic} if $\phi_{ci}$ and $\phi_{si}$ are conditional and simple invariants, and $\phi_{md}$ is monotonically decreasing for $\phi_{si}$.
The reason is that the characteristic function $\charfun{\phi}$ with $\charfun{\phi} = 1 \iff \phi$
and $\charfun{\phi} = 0 \iff \neg\phi$ of conditional invariants like $\phi_{ci}$ and $\phi_{si}$ is monotonically
increasing w.r.t.\ $\update{\alpha}$ and \eqref{eq:inverse-consecution} essentially requires that $\charfun{\phi_{md}}$
is monotonically decreasing w.r.t.\ $\update{\alpha}$.

\begin{example}[Invariants]\label{ex:invariants}
  For $\alpha_2$ from \Cref{ex:leading}, $y \geq 0$ is a simple
  invariant and $x \geq 0$ is monotonically decreasing for $y \geq 0$, as
  \[
    \begin{array}{llll}
      \forall x,y.\ y \geq 0 &\implies& y + 1 \geq 0 & \text{and}\\
      \forall x,y.\ y \geq 0 \land x - y \geq 0 &\implies& x \geq 0
    \end{array}
  \]
  are valid. Thus, $y \geq 0$ is also a conditional invariant.
  Note that it is not a standard invariant as there are program runs where $y \geq 0$ is violated when
  $\alpha_2$ is applied.
\end{example}

\smallskip

\subsection{Loop Acceleration}
\label{subsec:acceleration}

 The key
idea of loop acceleration for a \emph{simple loop}, i.e., a transition
$\alpha$ with $\head{\alpha} = \target{\alpha}$, is to generate
a new
transition $\overline{\alpha}$ that captures $\tv$ iterations of $\alpha$.
Here, $\tv$ is a fresh
variable whose value can be chosen non-deterministically. We first use
\emph{recurrence solving}
to compute closed forms for the values of the
program variables after a symbolic number of iterations, i.e., a closed form of
$\update{\alpha}^\tv = \underbrace{\update{\alpha} \circ \ldots \circ
  \update{\alpha}}_{\tv \text{ times}}$.
Then, as in \cite{underapprox15}, we exploit the
following observation: If $\guard{\alpha}$ holds after $\tv-1$ loop iterations and $\update{\alpha}(\guard{\alpha})$
implies $\guard{\alpha}$ (i.e., $\guard{\alpha}$ is monotonically decreasing),
then $\guard{\alpha}$ also holds after $\tv-2,\tv-3,\ldots,0$
iterations.
Thus, adding $\update{\alpha}^{\tv-1}(\guard{\alpha})$ to  $\guard{\overline{\alpha}}$
ensures that $\tv$ only takes feasible values: If
$\sigma$ satisfies $\update{\alpha}^{\tv-1}(\guard{\alpha})$, then $\alpha$ can be iterated
at least $\sigma(\tv)$ times.

However, $\update{\alpha}(\guard{\alpha}) \Longrightarrow \guard{\alpha}$ is rarely valid
  if $\guard{\alpha}$ contains
  invariants of $\alpha$.  Thus,
  our novel loop acceleration technique only requires monotonicity of $\guard{\alpha}$ instead.

\begin{theorem}[\accel]
  \label{thm:loop-acceleration}
  Let $\PP$ be well formed, let \(\alpha \in \PP\) be a simple loop with
  $\lhs{\alpha} = \ff(\vect{x})$,
  let $\tv \in \VV$ be
  fresh, and let $\mu$ be a substitution such that \(\mu(\vect{x}) = \update{\alpha}^{\tv}(\vect{x})\) holds for
  all $\tv > 0$.  Moreover, let $\guard{\alpha} = \phi_{ci} \land \phi_{si} \land \phi_{md}$ be monotonic.
  Finally, let $\dec_\tv = \{\tv \mapsto \tv-1\}$ and  $\overline{\PP} = \PP
  \cup\{ \overline{\alpha} \}$ where
  \[
    \overline{\alpha} = \boxed{\ff(\vect{x}) \tox{c}{}{} \ff(\mu(\vect{x}))
      \constr{\phi_{ci} \land \phi_{si} \land \dec_\tv(\mu(\phi_{md})) \land \tv\!>\!0}}\;.
  \]
  Then the processor \accel: \(\PP \mapsto \overline{\PP}\) is safe and
 sound.
\end{theorem}
\makeproof{thm:loop-acceleration}{
  We first show that for any valuation $\sigma$, $\sigma \models
  \guard{\overline{\alpha}}$ implies
  \begin{equation}
    \label{eq:proof-accel-guard-sat}
    \sigma \models \update{\alpha}^n(\guard{\alpha}) \text{ for all } 0 \leq n < \sigma(k).
  \end{equation}
  Let $\sigma$ be a valuation with $\sigma \models \guard{\overline{\alpha}}$. As $\phi_{si}$ is a simple
  invariant, $\phi_{si}$ implies $\update{\alpha}(\phi_{si})$. Since $\phi_{si} \subseteq
  \guard{\overline{\alpha}}$, we
  obtain
  \begin{equation}
    \label{eq:proof-accel-si}
    \sigma \models \update{\alpha}^n(\phi_{si}) \text{ for all } n \in \NN.
  \end{equation}

  As we also have $\dec_\tv(\mu(\phi_{md})) \subseteq \guard{\overline{\alpha}}$, we get
  \begin{equation}
    \label{eq:proof-accel-guard-md}
    \sigma \models \update{\alpha}^{\tv-1}(\phi_{md}),
  \end{equation}
  by definition of $\dec_\tv$ and $\mu$.

  We now use induction on $m$ to prove
  \begin{equation}
    \label{eq:proof-accel-md-induct}
    \sigma \models \update{\alpha}^{\tv - m}(\phi_{md}) \text{ for all } 1 \leq m \leq \sigma(\tv).
  \end{equation}
  The induction base $m = 1$ is immediate due to \eqref{eq:proof-accel-guard-md}. For the
  induction step, let $m > 1$. Due to the induction hypothesis, we obtain
   \begin{equation}
    \label{eq:k-m+1}
    \sigma \models \update{\alpha}^{\tv - m + 1}(\phi_{md}).
  \end{equation}
  By
  \eqref{eq:inverse-consecution}, we have
 \[
 \sigma \models  \update{\alpha}^{\tv -
      m}(\phi_{si}) \land \update{\alpha}^{\tv - m + 1}(\phi_{md})
    \implies \update{\alpha}^{\tv - m}(\phi_{md}).
    \]
  Hence,  \eqref{eq:k-m+1}  and \eqref{eq:proof-accel-si} imply
    \[
     \sigma \models  \update{\alpha}^{\tv - m}(\phi_{md}).
       \]
      This finishes the proof of \eqref{eq:proof-accel-md-induct}, which is equivalent to
  \begin{equation}
    \label{eq:proof-accel-md}
    \sigma \models \update{\alpha}^{n}(\phi_{md}) \text{ for all } 0 \leq n < \sigma(\tv).
  \end{equation}

  We now prove\footnote{Note that in contrast to the statement
    \eqref{eq:proof-accel-si} for simple invariants, $\sigma \models
    \update{\alpha}^{n}(\phi_{ci})$ does not necessarily hold for all $n \in \NN$. This is
    the reason why we distinguish between simple and conditional invariants and only use
    simple invariants in the premise of \eqref{eq:inverse-consecution}.}
  \begin{equation}
    \label{eq:proof-accel-ci}
    \sigma \models \update{\alpha}^{n}(\phi_{ci}) \text{ for all } 0 \leq n \leq \sigma(\tv)
  \end{equation}
  by induction on $n$.
      For the induction base ($n=0$) we have $\sigma \models \phi_{ci}$ since
  $\phi_{ci} \subseteq \guard{\overline{\alpha}}$.
  For the induction step, let $n>0$. Due to the induction hypothesis, we get
  \[
    \sigma \models \update{\alpha}^{n-1}(\phi_{ci}).
  \]
  Moreover, we have
  \[
    \sigma \models \update{\alpha}^{n-1}(\phi_{si})
  \]
  and
  \[
    \sigma \models \update{\alpha}^{n-1}(\phi_{md})
  \]
  due to \eqref{eq:proof-accel-si} and \eqref{eq:proof-accel-md}. As
  $\guard{\alpha} = \phi_{ci} \land \phi_{si} \land \phi_{md}$, we obtain
  \[
    \sigma \models \update{\alpha}^{n-1}(\guard{\alpha}).
  \]
  As $\phi_{ci}$ is a conditional invariant and $\phi_{ci} \subseteq \guard{\alpha}$, $\guard{\alpha}$
  also implies $\update{\alpha}(\phi_{ci})$. Thus, we get
  \[
    \sigma \models \update{\alpha}^{n}(\phi_{ci}).
  \]
  This finishes the proof of
  \eqref{eq:proof-accel-ci}.

  Since, again, $\guard{\alpha} = \phi_{ci} \land \phi_{si} \land \phi_{md}$, \eqref{eq:proof-accel-si},
  \eqref{eq:proof-accel-md}, and \eqref{eq:proof-accel-ci} together imply
  \[
    \sigma \models \update{\alpha}^{n}(\guard{\alpha}) \text{ for all } 0 \leq n < \sigma(\tv),
  \]
  which finishes the proof of \eqref{eq:proof-accel-guard-sat}.

  Thus, we have
  \[
    \sigma(\lhs{\overline{\alpha}}) = \ff(\sigma(\update{\alpha}^0(\vect{x}))) \tox{}{*}{\PP} \ff(\sigma(\update{\alpha}^{\sigma(k)}(\vect{x})))
  \]
  for any model $\sigma$ of $\guard{\overline{\alpha}}$, where the evaluation takes
  $\sigma(\tv) > 0$ steps.
   As
  \[
    \ff(\sigma(\update{\alpha}^{\sigma(k)}(\vect{x}))) = \ff(\sigma(\mu(\vect{x}))) = \sigma(\rhs{\overline{\alpha}})
  \]
  by the definition of $\mu$, this shows that
 any evaluation with $\overline{\alpha}$ can be
replaced by at least one evaluation step with $\alpha$. Thus, every non-terminating
evaluation with $\overline{\PP}$ can be transformed into a corresponding non-terminating
evaluation with $\PP$. Hence,
  the processor \accel is sound.

The processor is also safe, i.e., $\overline{\PP}$ is well formed. The reason is that by
the argumentation above, if $\sigma$ is a valuation with $\sigma \models
\guard{\overline{\alpha}}$, then
the configuration $\sigma(\lhs{\overline{\alpha}})$ can be evaluated to
$\sigma(\rhs{\overline{\alpha}})$ with the well-formed program $\PP$
and thus,  $\sigma(\rhs{\overline{\alpha}})$ must also be a configuration.}

So to construct $\rhs{\overline{\alpha}}$, we compute a closed form $\mu$ that expresses $\tv$ iterations
of the loop body as in
\cite{ijcar16,underapprox15}. To do so, one can use state-of-the-art recurrence solvers
like
\cite{purrs,mathematica,maple} to solve the
system of recurrence relations $\vect{x}^{(\tv + 1)} = \update{\alpha}(\vect{x}^{(\tv)})$ with the initial condition
$\vect{x}^{(1)} = \update{\alpha}(\vect{x})$.

To see why \accel is sound,
assume that $\guard{\overline{\alpha}}$
holds. As\footnote{In the following, we identify conjunctions and sets of inequations.}
$\phi_{si} \subseteq \guard{\overline{\alpha}}$
and $\phi_{si}$ implies $\update{\alpha}(\phi_{si})$ by \eqref{eq:consecution-simple}, we obtain
\begin{equation}
  \label{eq:proof-sketch-1}
  \update{\alpha}^n(\phi_{si}) \text{ for all } n \in \NN.
\end{equation}
Thus, as $\guard{\overline{\alpha}}$ contains
$\dec_\tv(\mu(\phi_{md})) = \update{\alpha}^{\tv-1}(\phi_{md})$ and $\phi_{si} \land
\update{\alpha}(\phi_{md})$ implies
$\phi_{md}$ by \eqref{eq:inverse-consecution}, we get
\begin{equation}
  \label{eq:proof-sketch-2}
  \update{\alpha}^{n}(\phi_{md}) \text{ for all } 0 \leq n < \tv.
\end{equation}
So  \eqref{eq:proof-sketch-1} and \eqref{eq:proof-sketch-2} imply
$\phi_{si} \land \phi_{md}$. As
 $\phi_{ci} \subseteq \guard{\overline{\alpha}}$ and
$\guard{\alpha} = \phi_{ci} \land \phi_{si} \land \phi_{md}$,
this means that $\guard{\alpha}$ holds as well.
As $\guard{\alpha}$ implies $\update{\alpha}(\phi_{ci})$
(since $\phi_{ci} \subseteq \guard{\alpha}$ and $\phi_{ci}$
is a conditional invariant), we obtain that $\update{\alpha}(\phi_{ci})$ holds.
Together with \eqref{eq:proof-sketch-1} and \eqref{eq:proof-sketch-2} this
means that $\update{\alpha}(\guard{\alpha})$ holds
(if $1 < \tv$). This in turn implies $\update{\alpha}^2(\phi_{ci})$, etc. Thus, we
get
\begin{equation}
  \label{eq:proof-sketch-3}
  \update{\alpha}^n(\phi_{ci}) \text{ for all } 0 \leq n \leq \tv.
\end{equation}
Due to \eqref{eq:proof-sketch-1} -- \eqref{eq:proof-sketch-3},
the constraint $\phi_{ci} \land \phi_{si} \land \dec_\tv(\mu(\phi_{md}))$ ensures that $\guard{\alpha} =
\phi_{ci} \land \phi_{si} \land \phi_{md}$ holds before the $1^{st}, \ldots, \tv^{th}$
iteration, as desired.
Hence, every evaluation with $\overline{\alpha}$ can be replaced by $\tv$ evaluation steps
with $\alpha$. Since $\guard{\overline{\alpha}}$ enforces $\tv > 0$, every non-terminating run with $\overline{\PP}$ can
therefore be transformed into a non-terminating run of $\PP$.

\begin{example}[\Cref{ex:leading} continued]
  \label{ex:acceleration}
  Consider the simple loop $\alpha_2$ of \Cref{ex:leading}. As $x \geq 0$ is not monotonic,
  \accel is not applicable. But if we strengthen the guard by adding the simple invariant $y \geq 0$, then
  $x \geq 0$ satisfies \eqref{eq:inverse-consecution}, cf.\ \Cref{ex:invariants}.
  Thus,
  we can apply \accel with
$\phi_{ci}: \; \true$,  $\phi_{si}:\; y \geq 0$, and
  $\phi_{md}: \; x \geq 0$.   \Cref{sec:inv} will show how
  to find simple invariants like $y \geq 0$.

  To compute a substitution $\mu$ that represents $k$ repeated updates,
  we solve the recurrence relations $y^{(\tv + 1)} = y^{(\tv)} + 1$ and $x^{(\tv+1)} = x^{(\tv)} -
  y^{(\tv)}$ with the initial conditions $x^{(1)} = x - y$ and $y^{(1)} = y + 1$, resulting in the solutions $y^{(\tv)} = y +
  \tv$ and $x^{(\tv)} = x-y \cdot \tv - \frac{1}{2} \cdot \tv^2 + \frac{1}{2} \cdot \tv$, i.e., $\mu = \{x \mapsto x-y
  \cdot \tv - \frac{1}{2} \cdot \tv^2 + \frac{1}{2} \cdot \tv, \; y \mapsto y +
  \tv\}$. Thus, we accelerate $\alpha_2$ to
  \[
    \overline{\alpha}_2:\quad \ff(x,y) \tox{\tv}{}{} \ff(\underbrace{x-y \cdot \tv - \tfrac{1}{2} \cdot \tv^2 +
      \tfrac{1}{2} \cdot \tv}_{\mu(x)}, \;
    \underbrace{y + \tv\vphantom{\tfrac{1}{2}}}_{\mu(y)}) \constr{\eta}
  \]
  for
  $
    \eta:\; \underbrace{y \geq 0}_{\phi_{si}} \land \dec_\tv(\mu(\phi_{md})) \land \tv > 0
  $
  where $\dec_\tv(\mu(\phi_{md}))$ is
  $
    x-y \cdot (\tv-1) - \tfrac{1}{2} \cdot (\tv-1)^2 + \tfrac{1}{2} \cdot (\tv - 1) \geq 0.
  $
\end{example}

\medskip

\subsection{Chaining}
\label{subsec:chaining}

\accel only applies to
simple loops.
 To transform loops with complex control
flow into simple loops and to eventually obtain simplified programs,
we use \emph{chaining}, a standard technique to combine two transitions
  $\ff(\ldots) \to
\fg(\ldots)$ and\ $\fg(\ldots) \to \fh(\ldots)$ to a new transition $\ff(\ldots) \to
\fh(\ldots)$  that captures the
effect of both transitions after each other.

\begin{theorem}[\chain]
  \label{thm:chaining}
  Let $\PP$ be well formed and let $\alpha, \beta \in \PP$ where $\target{\alpha} = \head{\beta}$, the argument lists of
  $\lhs{\alpha}$ and
  $\lhs{\beta}$
  are equal, and $\VV(\alpha) \cap \VV(\beta) = \VV(\lhs{\alpha})$.\footnote{Otherwise, one can rename variables without affecting the relation ${\to_\PP}$.} Let
    $\PP^\circ = \PP \cup
  \{\alpha \circ \beta\}$ with
  \[
    \alpha \circ \beta = \boxed{\lhs{\alpha} \tox[0pt]{\cost{\alpha} + \update{\alpha}(\cost{\beta})}{}{} \update{\alpha}(\rhs{\beta}) \constr{\guard{\alpha} \land \update{\alpha}(\guard{\beta})}}\;.
  \]
  Then the processor $\chain: \PP \mapsto \PP^\circ$ is safe and sound.
\end{theorem}
\makeproof{thm:chaining}{Let $\sigma$ be a model of $\guard{\alpha} \land
  \update{\alpha}(\guard{\beta})$
  with $\VV(\alpha) \cup \VV(\beta) \subseteq \dom(\sigma)$.
  Then $\sigma
  \models \guard{\alpha}$ implies $\sigma(\lhs{\alpha}) \to_\PP \sigma(\rhs{\alpha}) =
  \sigma(\update{\alpha}(\lhs{\beta}))$,
as   $\target{\alpha} = \head{\beta}$ and  the argument lists of $\lhs{\alpha}$
  and $\lhs{\beta}$ are equal.
  From $\sigma \models \update{\alpha}(\guard{\beta})$
and the fact that $\sigma \circ  \update{\alpha}$ is a valuation due to well-formedness of
$\{ \alpha\}$,
  we obtain $\sigma \circ
  \update{\alpha} \models \guard{\beta}$. Thus, we get $\sigma(\update{\alpha}(\lhs{\beta})) \to_\PP
  \sigma(\update{\alpha}(\rhs{\beta}))$, i.e., we obtain $\sigma(\lhs{\alpha \circ \beta}) = \sigma(\lhs{\alpha})
  \tox{}{*}{\PP} \sigma(\update{\alpha}(\rhs{\beta})) = \sigma(\rhs{\alpha \circ \beta})$,
  where the evaluation with $\PP$ takes two steps. This proves soundness of \chain. Well-formedness of
$\PP^\circ$ follows from the fact that the valuation $\sigma \circ  \update{\alpha}$
  satisfies $\guard{\beta}$ and $\{\beta\}$ is well formed.}

Chaining is not only useful to transform complex
into simple loops, but it can also be used to combine a simple loop $\alpha$
with itself in order to enable loop acceleration  and to obtain
better closed forms for $\update{\alpha}^{\tv}$.
For example, consider
  the following loop, where the sign of $x$ alternates:
\[
\alpha_{neg}: \quad  \ff(x,y) \to \ff(-x,y-1) \constr{y > x}
\]
The closed form $(-1)^\tv \cdot x$ for the value of $x$ after $\tv$ iterations involves exponentials even though
$x$ does not grow exponentially. This is disadvantageous, as our implementation relies
on SMT solving, but SMT solvers have limited support for non-polynomial arithmetic.
Moreover, \accel is not applicable, as
$y > x$ is non-monotonic. However,
  this can be resolved by
chaining $\alpha_{neg}$ with itself, which results in
\[
 \alpha_{neg} \circ \alpha_{neg}: \quad \ff(x,y) \tox{2}{}{} \ff(x,y-2) \constr{y > x \land y-1 > -x}.
\]
This transition can be accelerated to
\[
  \ff(x,y) \tox{2 \cdot \tv}{}{} \ff(x,y-2 \cdot \tv) \constr{\dec_\tv(\mu(\guard{\alpha_{neg} \circ \alpha_{neg}})) \land \tv > 0}
  \]
where $\dec_\tv(\mu(\guard{\alpha_{neg} \circ \alpha_{neg}}))$ is
\[
  y - 2 \cdot (\tv - 1) > x \land y - 2 \cdot (\tv - 1) - 1 > -x,
  \]
  i.e., the accelerated transition does not contain exponentials.

  So for simple loops $\alpha$ that \emph{alternate the sign of a variable}
  (i.e.,
where $\update{\alpha}(x) = c \cdot x + t$ for some $x \in \VV(\lhs{\alpha})$, $c < 0$,
and $t \in \AA$ with $x \notin \VV(t)$), we accelerate $\alpha
\circ \alpha$ instead of $\alpha$.

Chaining can also help to obtain simpler closed forms for
transitions where variables are set to constants.
For example, a closed form for the repeated update of the variable $z$
in
\[
 \alpha_{const}: \quad \ff(x,y,z) \to \ff(x-1,2,y) \constr{x > 0}
\]
is $0^{\tv-1} \cdot y + (1 - 0^{\tv-1}) \cdot 2$, which is again not polynomial.
However, chaining $\alpha_{const}$  with itself yields
\[
  \alpha_{const} \circ \alpha_{const}:\quad \ff(x,y,z) \to \ff(x-2,2,2) \constr{x > 1}
\]
(where we simplified the guard), which can be accelerated to
\[
  \ff(x,y,z) \to \ff(x - 2 \cdot \tv,2,2) \constr{x - 2 \cdot (\tv - 1) > 1 \land \tv > 0},
\]
i.e., the accelerated transition again only contains polynomials.

Finally, chaining can also make acceleration applicable to loops that permute arguments:
\[
\alpha_p:\quad  \ff(x,y) \to \ff(y-1,x-1) \constr{x > 0}
\]
While $\alpha_p$ violates the prerequisites of \accel,
\[
\alpha_{p} \circ \alpha_{p}:\quad   \ff(x,y) \to \ff(x-2,y-2) \constr{x > 0 \land y - 1 > 0}
\]
can be accelerated to:
\[
  \ff(x,y) \to \ff(x-2 \cdot \tv,y-2 \cdot \tv) \constr{\dec_\tv(\mu(\guard{\alpha_{p} \circ \alpha_{p}})) \land \tv > 0}
\]

So to handle simple loops $\alpha$ where some variables ``stabilize''
(i.e., $\update{\alpha}^n(z) \in \ZZ$ for some
$z \in \VV$ and some $n>1$, as in $\alpha_{const}$) or where arguments are permuted  (as in
$\alpha_p$), we re\-peat\-ed\-ly chain $\alpha$ with itself
 as long as this reduces the size of
 \begin{equation}
   \label{eq:set-reduce}
   \{x \in \VV(\lhs{\alpha}) \mid \VV(\update{\alpha}(x)) \neq \emptyset \land x \notin \VV(\update{\alpha}(x))\}.
 \end{equation}

 \medskip

\subsection{Proving Non-Termination}
\label{subsec:non-termination}

To detect non-terminating simple loops $\alpha$, we check whether $\guard{\alpha}$ itself is a simple invariant (i.e.,
whether the valuations that satisfy $\guard{\alpha}$ correspond to a \emph{recurrent
  set} of the relation ${\to_{\{\alpha\}}}$, cf.\ \cite{rupak08}).

\begin{theorem}[\nonterm]
  \label{thm:nonterm}
  Let $\PP$ be well formed and let \(\alpha \in \PP\) be a simple loop
  such that
  $\guard{\alpha}$ is a simple invariant.  More\-over,
  let $\PP^\infty = \PP \cup\{ \alpha^\infty \}$ where
  \[
    \alpha^\infty = \boxed{\textstyle \lhs{\alpha} \tox{\omega}{}{} \infty \constr{\guard{\alpha}}}\;.
  \]
  Then the processor \(\nonterm: \PP \mapsto \PP^\infty\) is safe and sound.
\end{theorem}
\makeproof{thm:nonterm}{Well-formedness of $\PP^\infty$ is trivial. For soundness, let
  $\sigma$ be a valuation with
  \[
    \sigma \models \guard{\alpha^\omega}.
  \]
  It suffices to prove that $\sigma(\lhs{\alpha})$ starts a non-terminating run with
  $\PP$.
  To this end, we prove $\sigma \models
  \update{\alpha}^n(\guard{\alpha})$ for all $n \in \NN$ by induction on $n$. The induction base is trivial, as
  $\guard{\alpha^\omega} = \guard{\alpha}$. For the induction step, let $n > 0$. The induction hypothesis implies
  $\sigma \models \update{\alpha}^{n-1}(\guard{\alpha})$. As $\guard{\alpha}$ is a simple invariant, it implies
  $\update{\alpha}(\guard{\alpha})$ and we obtain $\sigma \models \update{\alpha}^{n}(\guard{\alpha})$, which finishes
  the proof of the theorem.}

\begin{example}[\Cref{ex:leading} continued]
  \label{ex:nonterm}
    Clearly, $y > 0$ is not a simple invariant of $\alpha_4$ from \Cref{ex:leading}.
   But if we strengthen the guard by adding
  the simple invariant $x
  \leq 0$, then \nonterm is applicable as $y > 0 \land x \leq 0$ implies $y - x > 0
  \land x \leq 0$. Thus, we obtain
  \[
    \alpha_4^\infty: \quad \fg(x,y) \tox{\omega}{}{} \infty \constr{y > 0 \land x \leq 0}.
  \]
  Again, we will see how to deduce suitable simple
  invariants like $x \leq 0$ automatically in \Cref{sec:inv}.
\end{example}

In some cases,  chaining also helps to make \nonterm applicable. To see this, consider the simple loop
\[
 \alpha_{nt}: \quad \ff(x,y) \to \ff(0,y-x) \constr{y > 0}
 \]
where $y>0$ is no simple invariant. Chaining it with itself yields
\[
\alpha_{nt} \circ \alpha_{nt}: \quad \ff(x,y) \to \ff(0,y-x) \constr{y > 0 \land y - x > 0}.
\]
As $y > 0 \land y - x > 0 \implies y - x > 0 \land y - x - 0 > 0$ is valid, the prerequisites of \nonterm are
satisfied and we obtain
\[
  \ff(x,y) \to \infty \constr{y > 0 \land y - x > 0}.
\]

So in general, we try to apply \nonterm not only to a simple loop $\alpha$, but also to
$\alpha \circ \alpha$.
Apart from \nonterm, we also use SMT solving to check whether a loop has a fixpoint, which is a standard
technique  to prove non-termination.

\begin{theorem}[\fixedpoint]
  \label{thm:fixed-points}
  Let $\PP$ be well formed, let $\alpha \in \PP$ be a simple loop with $\lhs{\alpha} =
  \ff(\vect{x})$, and let
$\guard{\alpha} \land \vect{x} = \update{\alpha}(\vect{x})$ be satisfiable.
Let $\PP^{\bm{f\!p}} = \PP
  \cup \{\alpha^{\bm{f\!p}}\}$ where
  \[
    \alpha^{\bm{f\!p}} = \boxed{\lhs{\alpha} \to \infty \constr{\guard{\alpha} \land \vect{x} = \update{\alpha}(\vect{x})}}\;.
  \]
  Then the processor $\fixedpoint: \PP \mapsto \PP^{\bm{f\!p}}$ is safe and sound.
\end{theorem}
\makeproof{thm:fixed-points}{Well-formedness of $\PP^{\bm{f\!p}}$ is trivial. For soundness, let $\theta$ be a
  valuation with
  \[
    \theta \models \guard{\alpha^{\bf{f\!p}}}.
    \]
    Since $\guard{\alpha^{\bf{f\!p}}}$ implies
    $\vect{x} = \update{\alpha}(\vect{x})$, by induction on $n$ we obtain that
  \[
    \theta(\vect{x}) = \theta(\update{\alpha}^n(\vect{x})) \text{ holds for all
      for all $n \in \NN$.}
    \]
 Hence,
$\theta(\lhs{\alpha})$ starts a non-terminating run, because
 $\theta \circ
\update{\alpha}^n \models \guard{\alpha}$ for all $n \in \NN$.}

For example, $\{x \mapsto 0, y \mapsto 1\}$ is a fixpoint of $\alpha_{nt}$ which
satisfies $y > 0$ and $(x,y) = (0, y-x)$ (i.e., $x = 0 \land y = y-x$).

\Cref{alg:proving-non-term} shows a streamlined version of the strategy that we use to apply the presented processors.
It combines chaining, loop acceleration, and our non-termination
processors to transform
arbitrary programs into simplified programs.
For nested loops, the elimination starts with the inner loops.
Note that deleting transitions (Steps \ref{Step del 1}, \ref{Step del 2}, \ref{Step elimination alpha}, and \ref{Step delete 2}) is always sound in our
setting.

\begin{algorithm}
  \KwIn{A program $\PP$}
  \KwOut{A witness for non-termination of $\PP$ or $\bot$}
  \PWhile{$\PP$ is not simplified}{
    $\PP \leftarrow \{\alpha \in \PP \mid \head{\alpha} \text{ is reachable from } \start\}$\;
    \label{Step del 1}
    $\PP \leftarrow \{\alpha \in \PP \mid \guard{\alpha} \text{ is satisfiable}\};\quad \SSS \leftarrow \emptyset$\;
    \label{Step del 2}
    \PWhile{$\exists \alpha \in \PP.\ \alpha$ is a simple loop}{
      $\PP \leftarrow \PP \setminus \{\alpha\}$\;
      \label{Step elimination alpha}
          $\alpha \leftarrow \alpha \circ \alpha$ \Pif $\alpha$ alternates the sign of a
      variable \;
          $\alpha_{orig} \leftarrow \alpha$\;
          \lDo{it reduces the size of \eqref{eq:set-reduce}}{$\alpha \leftarrow \alpha \circ \alpha_{orig}$}
         \lIf{\nonterm applies to $\alpha$}{$\alpha \leftarrow \alpha^\infty$}\label{Step Nonterm 1}
         \lElsIf{\nonterm   applies to $\alpha \circ \alpha$}{$\alpha \leftarrow (\alpha \circ \alpha)^\infty$}
         \lElsIf{\fixedpoint applies to $\alpha$}{$\alpha \leftarrow \alpha^{\bm{f\!p}}$}
         \lElsIf{\accel applies to $\alpha$}{$\alpha \leftarrow \overline{\alpha}$}\label{Step Accel}
         \lElse{$\PP \leftarrow \PP \cup \deduceInvariant(\alpha)$ \label{Step invariant}}
         $\SSS \leftarrow \SSS \cup \{\beta \circ \alpha \mid \beta \in \PP, \;
         \head{\beta} \neq \target{\beta} = \head{\alpha}\}$\;
         \label{Step chaining 1}
       }
       $\PP \leftarrow \PP \cup \SSS$\;
       \If{$\exists \alpha, \beta \in \PP.\ \target{\alpha} = \head{\beta} = \ff$\label{Step if 1}}{
          $\PP \leftarrow \PP \cup \{\alpha \circ \beta \mid \alpha, \beta \in \PP, \;
         \target{\alpha} = \head{\beta} = \ff\}$\;
          \label{Step chaining 2}
          $\PP \leftarrow \{\alpha \in \PP \mid \ff \notin \{\head{\alpha}, \target{\alpha}\}\}$
        }\label{Step delete 2}
  }
  \lIf{$\exists \alpha \in \PP.\ \rhs{\alpha}\!=\!\infty \land \sigma\!\models\!\guard{\alpha}$\label{Step final}}
  {\Return{$\sigma(\lhs{\alpha})$}}
  \lElse{\Return{$\bot$}}
  \vspace{0.5em}
  \caption{Proving Non-Termination}
  \label{alg:proving-non-term}
\end{algorithm}

We present the algorithm $\deduceInvariant$ for Step \ref{Step invariant} in
\Cref{sec:inv}. It creates variants
of $\alpha$ by extending $\guard{\alpha}$ with suitable constraints
to make \accel
or \nonterm applicable.
Step \ref{Step chaining 1}
chains $\alpha$ with all
preceding transitions that are no simple loops.
Steps
 \ref{Step chaining 2} and \ref{Step delete 2} eliminate a function symbol
 via chaining.
 Note that
 \Cref{alg:proving-non-term} could have non-terminating runs, as it may add new transitions in Step \ref{Step
   invariant}. However, this turned out to be unproblematic in our experiments,
 cf.\ \Cref{sec:experiments}.

 \begin{example}[\Cref{ex:leading} finished]
   \label{ex:finish}
  After accelerating $\alpha_2$ in Step \ref{Step Accel} (see \Cref{ex:acceleration}), \Cref{alg:proving-non-term}
  computes
  \[
    \alpha_1 \circ \overline{\alpha}_2:\quad \start(x,y) \to \ff(x-y \cdot \tv - \tfrac{1}{2} \cdot \tv^2 + \tfrac{1}{2} \cdot \tv, y + \tv) \hfill \constr{\eta}
  \]
  in Step \ref{Step chaining 1} where $\eta$ is
  \[
    y \geq 0 \land x-y \cdot (\tv-1) - \tfrac{1}{2} \cdot (\tv-1)^2 + \tfrac{1}{2} \cdot (\tv - 1) \geq 0 \land \tv > 0.
  \]
  Next, it applies \nonterm to $\alpha_4$ in Step \ref{Step Nonterm 1} (see \Cref{ex:nonterm}) and chains the
  resulting transition with $\alpha_3$ in Step \ref{Step chaining 1}, which yields
  \[
    \alpha_3 \circ \alpha_4^\infty:\quad \ff(x,y) \to \infty \hfill \constr{x < 0 \land y > 0}.
  \]
  Then, it
  chains $\alpha_1 \circ \overline{\alpha}_2$ and $\alpha_3 \circ \alpha_4^\infty$ in Step \ref{Step chaining 2},
  resulting in
  \[
    \alpha_1 \circ \overline{\alpha}_2 \circ \alpha_3 \circ \alpha_4^\infty: \quad \start(x,y) \tox{}{}{} \infty \constr{\psi}
  \]
  where $\psi$ is
  $\eta \land \underbrace{x-y \cdot \tv - \tfrac{1}{2} \cdot \tv^2 +
      \tfrac{1}{2} \cdot \tv < 0}_{\update{\alpha_1 \circ \overline{\alpha}_2}(x < 0)} \land \underbrace{y + \tv > 0}_{\update{\alpha_1 \circ \overline{\alpha}_2}(y > 0)}$.

  To prove non-termination, we have to show satisfiability of $\psi$. As $\sigma \models
  \psi$ for $\sigma = \{x \mapsto 0, y \mapsto 0, \tv \mapsto 2\}$, the configuration $\sigma(\start(x,y)) = \start(0,0)$  witnesses
  non-termination of \Cref{ex:leading}.
\end{example}

So
loop acceleration introduces a new variable $\tv$ for the number of loop
  unrollings. Later, $\tv$ is instantiated when
  search\-ing for models of the guards of the simplified transitions
 which result
 from repeated acceleration and chaining.
In \Cref{ex:finish}, when inferring a model for the guard of $\alpha_1 \circ
\overline{\alpha}_2 \circ \alpha_3 \circ \alpha_4^\infty$,
  the instantiation $\tv \mapsto 2$
  means that $\alpha_2$ is applied twice in the corresponding non-terminating run of the original
  program.

\section{Deducing Simple Invariants}
\label{sec:inv}

In
\Cref{sec:simplify}, we have seen that we sometimes need to deduce suitable
simple invariants to apply
our novel loop acceleration technique or to prove non-termination. Soundness of adding constraints to
transitions is ensured by the following processor.

\begin{theorem}[\strengthen \cite{ijcar16}]
  \label{thm:strengthening}
  Let $\PP$ be well formed, let $\alpha \in \PP$, let $\phi$ be a constraint,
  and let $\PP^\bullet = \PP \cup \{\alpha^\bullet\}$ where
  \[
    \alpha^{\bullet} = \boxed{\lhs{\alpha} \to \rhs{\alpha} \constr{\guard{\alpha} \land \phi}}\;.
  \]
  Then the processor $\strengthen: \PP \mapsto \PP^\bullet$ is safe and sound.
\end{theorem}
\makeproof{thm:strengthening}{The theorem trivially holds, as
  \[
    \sigma(\lhs{\alpha^\bullet}) \to_{\{\alpha^{\bullet}\}} \sigma(\rhs{\alpha^\bullet})
  \]
  clearly implies
  \[
    \sigma(\lhs{\alpha^\bullet}) \to_{\{\alpha\}} \sigma(\rhs{\alpha^\bullet}).
  \]}

The challenge is to find constraints $\phi$
that help to prove non-termination.
We now explain how to automatically synthesize suitable simple
invariants to strengthen a simple
  loop $\alpha$, cf.\
  Contribution \ref{it:inv} from \Cref{sec:introduction}.
    Our approach iteratively generates simple invariants such that larger and larger
parts of $\guard{\alpha}$ become monotonic.
To this end, it constructs
arithmetic formulas
and uses constraint solvers to instantiate their free variables (or
\emph{parameters}) such that they become valid. This results in simple invariants that are suitable for
strengthening. Eventually,  our technique either fails to synthesize further invariants
or the whole guard becomes monotonic, so that  we can apply \accel or even
\nonterm
(if $\alpha$'s guard is a simple  invariant).

To synthesize simple invariants,
we first compute a maximal subset
$\phi_i$ of $\guard{\alpha}$
such that $\phi_{i}$ is a conditional
invariant.  However, to apply \accel, not all constraints of
$\guard{\alpha}$ need to be conditional invariants, as long as the remaining constraints are monotonically decreasing.
Hence, we next compute a maximal subset $\phi_{si}$ of $\phi_i$ such that $\phi_{si}$ is
a simple invariant. Then we can determine a maximal subset $\phi_{md}$ of $\guard{\alpha} \setminus\, \phi_i$ which is monotonically decreasing for $\phi_{si}$.

\subsection{Generating New Invariants}\label{subsec:generating-conditional-invariants}
Let the set of \emph{parameters} $\Params \subset \VV$ be countably infinite and disjoint from the program variables $\VV(\PP)$.
Moreover, let
$\phi_{nm} = \guard{\alpha} \setminus\; (\phi_i \cup \phi_{md})$, i.e., $\phi_{nm}$ causes
\underline{n}on-\underline{m}onotonicity of $\guard{\alpha}$.  For each inequation
  $\rho \in \phi_{nm}$, we construct a linear template $\temp{\rho}$
over the \emph{relevant} variables $\VV_\rho$ of $\rho$, i.e., $\VV_\rho$ is the smallest set such that $\VV(\rho) \subseteq
\VV_\rho$, $\VV(\rho') \cap \VV_\rho \neq \emptyset$ implies $\VV(\rho') \subseteq \VV_\rho$ for each $\rho' \in
\guard{\alpha}$, and $x \in \VV_\rho$ implies $\VV(\update{\alpha}(x)) \subseteq \VV_\rho$. So $\temp{\rho}$
has the form $\sum_{x \in \VV_\rho} c_x \cdot x \geq c$ where $\{c_x \mid x \in \VV_\rho\} \cup \{c\} \subset \Params$.

For $\alpha_4$ from \Cref{ex:leading}, we obtain $\phi_i = \emptyset$, $\phi_{md} = \emptyset$, and $\phi_{nm} = \{y > 0\}$. As $x \in
\update{\alpha_4}(y)$, we have $\VV_{y > 0} = \{x,y\}$. Hence, $\temp{y>0}$ is $c_x \cdot x + c_y \cdot y \geq c$ where $c_x,c_y,c \in \Params$.

To find a valuation of the parameters such that all templates can be added to $\phi_{si}$ without violating
the definition of simple invariants, we enforce \eqref{eq:consecution-simple} for
$\phi_{si} \land \bigwedge_{\rho \in \phi_{nm}} \temp{\rho}$ by requiring
\begin{equation}
  \tag{$\tau$-si}
  \label{eq:inv-template-invariant}
  \forall \VV(\PP).\
  \phi_{si} \land \bigwedge_{\rho \in \phi_{nm}} \temp{\rho} \implies \bigwedge_{\rho \in \phi_{nm}} \update{\alpha}(\temp{\rho}).
\end{equation}
So for $\alpha_4$, we search for a valuation of  $c_x, c_y,$ and $c$ that satisfies
\begin{equation}
  \tag{$\tau$-si-$\alpha_4$}
  \label{eq:inv-template-invariant-ex}
  \forall x,y.\
  c_x \cdot x + c_y \cdot y \geq c \implies c_x \cdot x + c_y \cdot (y-x) \geq c.
\end{equation}

\subsection{Improving Towards Monotonicity}
By construction, the constraint $\phi_i
\land \phi_{md}$ is monotonic. Furthermore,
\eqref{eq:inv-template-invariant} ensures that $\phi_{si} \land \bigwedge_{\rho \in \phi_{nm}}
\temp{\rho}$ is a simple invariant, i.e., we know that
\begin{equation}
  \label{eq:mon}
  \phi_i  \land \bigwedge_{\rho \in \phi_{nm}} \temp{\rho} \land \phi_{md}
\end{equation}
is monotonic.
Eventually, our goal is to turn $\guard{\alpha}$ into a simple invariant and apply
\nonterm
  or to make it monotonic and apply
\accel.
  To progress towards this goal incrementally,
  we ensure that we can add at least one $\rho \in \phi_{nm}$ to
  \eqref{eq:mon} without violating
  monotonicity.
  To this end, we enforce that \eqref{eq:consecution} or
  \eqref{eq:inverse-consecution} holds for some $\rho \in \phi_{nm}$ by requiring:
{
  \begin{align}
    \tag{some-ci}
    \label{eq:inv-psi-invariant}
    \bigvee_{\mathclap{\rho \in \phi_{nm}}}\
    & \forall \VV_\PP.\ \guard{\alpha} \land \bigwedge_{\mathclap{\xi \in \phi_{nm}}} \temp{\xi} \implies \update{\alpha}(\rho) \quad \text{or}\\
    \tag{some-md}
    \label{eq:inv-psi-decreasing}
    \bigvee_{\mathclap{\rho \in \phi_{nm}}}\
    & \forall \VV_\PP.\  \phi_{si} \land \bigwedge_{\mathclap{\xi \in \phi_{nm}}}
    \temp{\xi} \land \update{\alpha}(\phi_{md} \land \rho)
     \implies \rho
  \end{align}
}Note that \eqref{eq:inv-psi-invariant} can also help to apply \nonterm as $\guard{\alpha}$ is
a conditional invariant iff it is a simple invariant.

\subsection{Maximizing the Improvement}
It is clearly advantageous to instantiate the parameters in such a way that as many inequations from $\phi_{nm}$ as
possible can be added to \eqref{eq:mon} without violating monotonicity. Hence, we require
\begin{align}
  \tag{$\rho$-ci}
  \label{eq:inv-soft}
  & \forall \VV(\PP).\ \guard{\alpha} \land \bigwedge_{\mathclap{\xi \in \phi_{nm}}}
  \temp{\xi} \implies \update{\alpha}(\rho) \quad \text{or} \\
  \tag{$\rho$-md}
  \label{eq:inv-soft-decreasing}
  & \forall \VV(\PP).\ \phi_{si} \land \bigwedge_{\mathclap{\xi \in \phi_{nm}}}
    \temp{\xi} \land \update{\alpha}(\phi_{md} \land \rho)
    \implies \rho
\end{align}
for as
many $\rho \in \phi_{nm}$ as possible, i.e., each $\rho \in \phi_{nm}$ gives rise to a \emph{soft requirement}
$\eqref{eq:inv-soft} \lor \eqref{eq:inv-soft-decreasing}$. Later, soft requirements will be associated with
weights. We then try to maximize the weight of all valid soft requirements,
but some of them may be violated. However, all \emph{hard requirements} like \eqref{eq:inv-template-invariant}
and $\eqref{eq:inv-psi-invariant} \vee \eqref{eq:inv-psi-decreasing}$ must hold.

For $\alpha_4$, $\phi_{nm}$ is a singleton set and hence \eqref{eq:inv-psi-invariant} and \eqref{eq:inv-soft} resp.\
\eqref{eq:inv-psi-decreasing} and \eqref{eq:inv-soft-decreasing} coincide for $\rho:\; y > 0$.
{
  \begin{align}
    &\begin{split}
      &\hspace{-1.6em} \eqref{eq:inv-psi-invariant} / \eqref{eq:inv-soft}:\\
      &\hspace{-1.6em}\forall x,y.\ y > 0 \land c_x \cdot x + c_y \cdot y \geq c \implies y - x > 0
    \end{split}
        \label{eq:inv-psi-invariant-ex}
        \tag{$\rho$-ci-$\alpha_4$}
    \\
    &\begin{split}
      &\hspace{-1.6em} \eqref{eq:inv-psi-decreasing} / \eqref{eq:inv-soft-decreasing}:\\
      &\hspace{-1.6em}\forall x,y.\  c_x \cdot x + c_y \cdot y \geq c \land y-x > 0   \implies y > 0
    \end{split}
        \tag{$\rho$-md-$\alpha_4$}
  \end{align}
}

\subsection{Preferring Local Invariants}
\label{sec:local-invariants}
If we strengthen a transition $\alpha$ with an inequation
$\xi$, then the case $\neg\xi$ is not covered by
the resulting transition. So we split $\alpha$ relative to $\xi$, i.e., we
also strengthen $\alpha$ with $\neg\xi$.
  However, this increases the size of the
  program. Thus, we try to deduce standard invariants whenever possible, i.e.,
we try to deduce constraints $\xi$ that are valid whenever $\alpha$ is applied in a program run so that
the case $\neg\xi$ is irrelevant. To
detect such invariants in a mod\-u\-lar way,
we only consider \emph{local
  invariants}, i.e., constraints
whose invariance can be proven by reasoning about $\alpha$
and all tran\-si\-tions $\beta$ with $\target{\beta} = \head{\alpha}$, whereas all other
transitions are ignored.
A similar idea
is also used in
\cite{larraz14}  to synthesize invariants.

\begin{definition}[Local Invariants]
  \label{def:local-invariants}
  Let $\alpha \in \PP$. If $\phi_{li}$ is a con\-di\-tion\-al invariant of $\alpha$ and
for all $\beta \in \PP \setminus \{\alpha\}$ with $\target{\beta} = \head{\alpha}$,
  \begin{equation}
    \tag{li}
    \label{eq:initiation}
    \forall \VV(\PP).\
   \guard{\beta} \land \update{\beta}(\guard{\alpha}) \implies \update{\beta}(\phi_{li})
  \end{equation}
  is valid, then $\phi_{li}$ is a \emph{\underline{l}ocal \underline{i}nvariant} of $\alpha$.
\end{definition}

\Cref{def:local-invariants} requires that whenever $\beta$ can be applied ($\guard{\beta}$ in the premise
of \eqref{eq:initiation}) and $\alpha$ can be applied afterwards ($\target{\beta} =
\head{\alpha}$ and
$\update{\beta}(\guard{\alpha})$ in the premise of \eqref{eq:initiation}), then $\phi_{li}$
must hold after applying $\beta$ (which is the conclusion
of \eqref{eq:initiation}).

So for $\alpha_4$, $x \leq 0$ is clearly a simple invariant, as $\alpha_4$
does not update $x$. Moreover, the guard $x < 0$ of $\alpha_3$ (which is the only other transition whose destination is
$\head{\alpha_4}$) implies $x \leq 0$. Thus, $x \leq 0$ is a local invariant of $\alpha_4$.

To guide the search towards local
invariants, we add a soft requirement corresponding to \eqref{eq:initiation} for each $\rho \in \phi_{nm}$:
{
  \begin{equation}
    \tag{$\temp{\rho}$-li}
    \label{eq:inv-standard}
    \bigwedge_{\mathclap{\substack{\beta \in \PP \setminus \{\alpha\} \\ \target{\beta} =
          \head{\alpha}}}} \forall \VV(\PP).\
    \guard{\beta} \land \update{\beta}(\guard{\alpha}) \implies \update{\beta}(\temp{\rho})
  \end{equation}
}So
 for $\rho: \, y > 0$ in our example, due to transition $\alpha_3$
 we get:
\begin{equation}
  \tag{$\temp{\rho}$-li-$\alpha_4$}
  \label{eq:inv-standard-ex}
  \forall x,y.\ x < 0 \land y > 0 \implies c_x \cdot x + c_y \cdot y \geq c
\end{equation}

\smallskip

\subsection{Excluding Inapplicable Transitions}
So far we do not exclude
solutions that result in
inapplicable transitions. To solve this problem, we add the hard requirement
\begin{equation}
  \tag{sat}
  \label{eq:inv-init-sat}
  \bigvee_{\mathclap{\substack{\beta \in \PP \setminus \{\alpha\} \\ \target{\beta} = \head{\alpha}}}} \exists \VV(\PP).\
  \guard{\beta} \land \update{\beta}(\guard{\alpha}) \land \bigwedge_{\mathclap{\rho \in \phi_{nm}}}\update{\beta}(\temp{\rho}).
\end{equation}
So we require that there is a transition $\beta$ with $\target{\beta} = \head{\alpha}$ (due to the
leading $\bigvee \ldots$) and a valuation (due to the existential quantifier) such that $\beta$ is applicable (due to
$\guard{\beta}$) and $\alpha$ is applicable afterwards (due to $\update{\beta}(\guard{\alpha}) \land \bigwedge_{\rho \in
  \phi_{nm}}\update{\beta}(\temp{\rho})$, as we will strengthen $\alpha$'s guard with $\bigwedge_{\rho \in
  \phi_{nm}}\temp{\rho}$ after instantiating the parameters in the templates).  Thus, for
$\alpha_4$ we
require
\begin{equation}
  \tag{sat-$\alpha_4$}
  \label{eq:inv-init-sat-ex}
  \exists x,y.\ x < 0 \land y > 0 \land c_x \cdot x + c_y \cdot y \geq c
\end{equation}
due to the transition $\alpha_3$.

\Cref{alg:proving-non-term} essentially compresses each path through a
multi-path loop (e.g., a loop whose body contains case analyses)
into a simple loop via chaining
in order to apply \nonterm, \fixedpoint, or \accel afterwards.
So our technique tends to generate many
simple loops for function symbols that correspond to entry points of
multi-path loops.
Therefore, \eqref{eq:inv-standard} and \eqref{eq:inv-init-sat} can result in large formulas,
which leads to performance issues. Hence, our implementation only considers transitions $\beta$ with
$\head{\beta} \neq \target{\beta}$ when constructing
\eqref{eq:inv-standard} and \eqref{eq:inv-init-sat}.
Note that this is uncritical
for correctness, as the technique presented in the current section is only a heuristic
to generate constraints to be added via
\strengthen (which is always sound).

\medskip

\subsection{Preferring \nonterm}
Finally, we prefer simple invariants that allow us to apply
\nonterm, our main technique to prove \underline{n}on-\underline{t}ermination. To this end,
we add a soft requirement to prefer solutions where the guard of the resulting strengthened transition is a conditional
invariant whenever $\phi_{md}$ is empty:
\begin{equation}
  \tag{nt}
  \label{eq:inv-nt}
  \forall \VV(\PP).\ \guard{\alpha} \land \bigwedge_{\rho \in \phi_{nm}} \temp{\rho} \implies \bigwedge_{\rho \in \phi_{nm}} \update{\alpha}(\rho)
\end{equation}
In our example, \eqref{eq:inv-nt} equals \eqref{eq:inv-psi-invariant-ex} as $\phi_{nm}$ is a singleton
set.

\medskip

\subsection{Algorithm for Inferring Simple Invariants}\label{subsec:inv-alg}
\Cref{alg:inv-non-term} summarizes our approach to deduce simple invariants.
Here, the $i^{th}$ entry of the weight
vector $\vect{w}$ corresponds to the weight of the $i^{th}$ soft requirement $\chi_i$ and  $\solve(\zeta,
\vect{\chi}, \vect{w})$ searches an instantiation $\sigma$ of the parameters such that $\sigma \models \zeta$ and
$\sum_{\substack{1 \leq i \leq |\vect{\chi}| \\ \sigma \models \chi_i}} w_i$ is maximized.
We explain how to implement $\solve$ in \Cref{subsec:greedy}.
The weights are chosen in such a way that a solution $\sigma$ is preferred over
$\sigma'$ if $\sigma$ turns more templates $\temp{\rho}$  into local invariants than
$\sigma'$: The weight $m+2 =
|\phi_{nm}|+2$ of the formulas resulting from \eqref{eq:inv-standard} (where $|\phi_{nm}|$ is the number of inequations
in $\phi_{nm}$) ensures that each formula from $\eqref{eq:inv-standard}$ has
a higher weight than the sum of all other soft requirements $\eqref{eq:inv-soft} \lor \eqref{eq:inv-soft-decreasing}$ and \eqref{eq:inv-nt}.

\begin{algorithm}
  \KwIn{A simple loop $\alpha$}
  \KwOut{A set of strengthened variants of $\alpha$}
  \lIf{$\phi_{nm} = \emptyset$}{\Return{$\emptyset$}}
  \lElse{$res \leftarrow \emptyset$}
  \PWhile{$\phi_{nm} \neq \emptyset$}{
    \label{Step while guard}
    $i \leftarrow 0; \  m \leftarrow |\phi_{nm}|$\;
    \PFor{$\rho \in \phi_{nm}$}{
      $i \leftarrow i + 1$\;
      $\chi_i \leftarrow \eqref{eq:inv-standard}; \quad w_i \leftarrow m+2$\;
      $\chi_{i+m} \leftarrow \eqref{eq:inv-soft} \lor \eqref{eq:inv-soft-decreasing}; \quad w_{i+m} \leftarrow 1$
    }
    \lIf{$\phi_{md} = \emptyset$}{$\chi_{i+m+1} \leftarrow \eqref{eq:inv-nt}; \quad w_{i+m+1} \leftarrow 1$}
    $\zeta \leftarrow \eqref{eq:inv-template-invariant} \land (\eqref{eq:inv-psi-invariant} \lor \eqref{eq:inv-psi-decreasing}) \land \eqref{eq:inv-init-sat}$\;
    $\sigma \leftarrow \solve(\zeta, \vect{\chi}, \vect{w})$\;
    \Return{$res$} \Pif $\solve$ failed\;
    \PFor{$\rho \in \phi_{nm}$ where $\sigma(\temp{\rho})$ is not a local invariant}{
      $res \leftarrow res \cup \{\lhs{\alpha} \to \rhs{\alpha}
        \constr{\guard{\alpha}\!\land\!\neg \sigma(\temp{\rho})}\}$
    }
    $\guard{\alpha} \leftarrow \guard{\alpha} \land \bigwedge_{\rho \in \phi_{nm}}
    \sigma(\temp{\rho})$
    \label{Step guard update}
  }
  \Return{$\{\alpha\} \cup res$}
  \vspace{0.5em}
  \caption{$\deduceInvariant$}
  \label{alg:inv-non-term}
\end{algorithm}

Note that Step \ref{Step guard update} updates $\guard{\alpha}$ in each iteration and
$\phi_{nm}$ is recomputed before checking the condition of the \textbf{while}-loop in Step
\ref{Step while guard}.
\Cref{alg:inv-non-term} terminates: $|\phi_{nm}|$ decreases in every iteration due to the hard
 requirement $\eqref{eq:inv-psi-invariant} \lor \eqref{eq:inv-psi-decreasing}$, which ensures that some $\rho \in
 \phi_{nm}$ becomes part of $\phi_i$ or $\phi_{md}$. Moreover, the hard requirement \eqref{eq:inv-template-invariant}
 ensures that each $\sigma(\temp{\rho})$ becomes part of $\phi_{si}$, so \Cref{alg:inv-non-term} never adds
 elements to $\phi_{nm}$.

In our example, \eqref{eq:inv-template-invariant-ex}, \eqref{eq:inv-psi-invariant-ex}, \eqref{eq:inv-standard-ex}, and
\eqref{eq:inv-init-sat-ex} are valid if $c_x = -1$ and $c_y = c = 0$. Hence, \Cref{alg:inv-non-term} suc\-cess\-ful\-ly
generates the local invariant $-x \geq 0$, i.e.,
$x \leq 0$. Afterwards, we can apply \nonterm to the strengthened loop as
in \Cref{ex:nonterm}.

\begin{example}[Deducing Simple Invariants for $\alpha_2$]
  Reconsider the simple loop $\alpha_2$ from \Cref{ex:leading}, where $\phi_i = \phi_{md} = \emptyset$ and
  $\phi_{nm} = \{x \geq 0\}$ as $\alpha_2$'s guard $x \geq 0$ is not monotonic.  Here, $\temp{x \geq 0}$ is $c_x \cdot x
  + c_y \cdot y \geq c$ as $y \in \VV(\update{\alpha}(x))$. So
  \eqref{eq:inv-template-invariant} becomes
  \begin{multline}
    \tag{$\tau$-si-$\alpha_2$}
    \label{eq:inv-template-invariant-decreasing-ex}
    \forall x,y.\
    c_x \cdot x + c_y \cdot y \geq c\\
    \implies c_x \cdot (x-y) + c_y \cdot (y+1) \geq c.
  \end{multline}
  Again, $\eqref{eq:inv-psi-invariant} \lor \eqref{eq:inv-psi-decreasing}$ coincides with $\eqref{eq:inv-soft} \lor
  \eqref{eq:inv-soft-decreasing}$ for $\rho: \, x \geq 0$.\\[-0.5em]
  \resizebox{0.495\textwidth}{!}{
    \hspace{-1em}
    \begin{minipage}{0.52\textwidth}
    \begin{align}
      \tag{$\rho$-ci-$\alpha_2$}
      \label{eq:inv-increasing-ex}
      &\forall x,y.\ x \geq 0  \land c_x \cdot x + c_y \cdot y \geq c \implies x-y \geq 0 \hspace{0.5em} \lor\\
      \tag{$\rho$-md-$\alpha_2$}
      \label{eq:inv-decreasing-ex}
      &\forall x,y.\  c_x \cdot x + c_y \cdot y \geq c \land x - y \geq 0 \implies x \geq 0
    \end{align}
    \end{minipage}
  }\\[0.5em]
  Next,
  \eqref{eq:inv-standard} gives rise to the requirement
  \begin{equation}
    \tag{$\tau_{\rho}$-li-$\alpha_2$}
    \label{eq:inv-standard-decreasing-ex}
    \forall x,y.\ x \geq 0 \implies c_x \cdot x + c_y \cdot y \geq c.
  \end{equation}
  Moreover, \eqref{eq:inv-init-sat} becomes
  \begin{equation}
    \tag{sat-$\alpha_2$}
    \label{eq:inv-init-sat-decreasing-ex}
    \exists x,y.\ x \geq 0 \land c_x \cdot x + c_y \cdot y \geq c.
  \end{equation}
  Finally, \eqref{eq:inv-nt} equals \eqref{eq:inv-increasing-ex}.  Thus, the hard requirement $\zeta$ is
  \[
    \eqref{eq:inv-template-invariant-decreasing-ex} \land (\eqref{eq:inv-increasing-ex} \lor \eqref{eq:inv-decreasing-ex}) \land \eqref{eq:inv-init-sat-decreasing-ex}.
  \]
  The soft requirements are  \eqref{eq:inv-standard-decreasing-ex},
  $\eqref{eq:inv-increasing-ex} \lor \eqref{eq:inv-decreasing-ex}$,
  and \eqref{eq:inv-increasing-ex} with weights $3$, $1$, and $1$, respectively.
  The valuation $\sigma = \{c_x \mapsto 0, c_y \mapsto 1, c \mapsto 0\}$
  satisfies $\zeta$ and $\eqref{eq:inv-increasing-ex} \lor \eqref{eq:inv-decreasing-ex}$,
  but not the other soft constraints. As $\zeta \land \eqref{eq:inv-standard-decreasing-ex}$ and
  $\zeta \land \eqref{eq:inv-increasing-ex}$ are unsatisfiable, $\sigma$ is an optimal solution.
  It corresponds to the simple invariant $y \geq 0$. After deducing it, the
  strengthened transition can be accelerated as in
  \Cref{ex:acceleration}.
\end{example}

\subsection{Greedy Algorithm for Max-SMT Solving}\label{subsec:greedy}

We now
explain how to implement the function $\solve$ that is called in
\Cref{alg:inv-non-term} to
instantiate the parameters in the formulas. Our implementation is restricted to the case that these formulas are
linear w.r.t.\ the program variables $\VV(\PP)$. Then the universally quantified variables
can be eliminated by applying Farkas' Lemma \cite{bradley05,podryb}.
In this way, we obtain a \emph{Max-SMT} obligation
over the theory of non-linear
  integer\footnote{Note that rational constants can be eliminated by multiplying with the least common multiple.} arithmetic. While there exist powerful Max-SMT solvers
  \cite{z3,barcelogic,yices},
  we use a
straightforward greedy algorithm based on incremental SMT solving.
This approach turned out to be be more efficient than sophisticated Max-SMT techniques in our
setting, presumably as it does not aim to find provably optimal solutions.

\section{Experiments}
\label{sec:experiments}

We implemented our approach in our tool \tool{LoAT}
\cite{ijcar16}
which uses the recurrence solver \tool{PURRS} \cite{purrs} and the SMT solver \tool{Z3} \cite{z3}.
It supports the SMT-LIB input format \cite{smtlib-format} and the native formats of the
tools \tool{KoAT} \cite{koat}
and \tool{T2} \cite{t2-tool}.
We evaluated it on the benchmark suite from the \emph{Termination and Complexity Competition} (\tool{TermComp} \cite{termcomp})
consisting of 1222 programs (\tool{TPDB} \cite{tpdb}, category \emph{Termination of Integer Transition Systems}).
All experiments
were executed on \tool{StarExec} \cite{starexec} with a timeout of 60 seconds per example.

We first compared our new implementation with our technique to prove \underline{l}ower
complexity \underline{b}ounds of integer programs
from \cite{ijcar16} (\tool{LoAT~LB}), which can also deduce non-termination as a byproduct. \tool{LoAT~LB} proves non-termination
in 390 cases,
whereas the new version of \tool{LoAT} succeeds for 462 examples.

Then, we compared \tool{LoAT} with two
  state-of-the-art termina\-tion analyzers for integer programs:
\tool{VeryMax} \cite{borralleras17,larraz14} (resp.\ its predecessor \tool{CppInv}) won the
category \emph{Termination of Integer Transition Systems} at
\tool{TermComp} in 2014 and 2016 -- 2019. \tool{T2} was the winner in
2015. We
  also tested with our tool \tool{AProVE} \cite{tool-jar}, but excluded it as it uses a similar
  approach like \tool{T2}, but finds fewer non-termination proofs.
    The
      remaining participants of
the respective category of \tool{TermComp}, \tool{Ctrl} \cite{kop15} and \tool{iRankFinder}
\cite{amram18, irankWST},
cannot prove non-termination.\footnote{\tool{iRankFinder} can prove non-termination of simple loops \cite{amram18},
  but according to its authors it cannot yet
  check reachability of diverging configurations.}
We used the \tool{TermComp} '19 version of \tool{VeryMax} and the
\tool{TermComp} '17 version of \tool{T2} (as \tool{T2} has not been developed further since 2017).
Our experiments did not reveal any con\-flicts, i.e., there is no example where
one tool proved termi\-nation and another  proved non-termination.

\noindent
\begin{tabularx}{0.49\textwidth}{|c||Y|Y|Y|Y|Y|}
  \hline
  & \tool{LoAT} & \tool{T2} & \tool{VeryMax} \\ \hline \hline
  NO          &         462 &       420 &            392 \\ \hline
  YES         &           0 &       607 &            623 \\ \hline
  MAYBE       &         760 &       195 &            207 \\ \hline
  Unique NO   &          22 &         9 &             23 \\ \hline
  Avg.\ time  &        8.3s &      8.8s &          13.8s \\ \hline
\end{tabularx}

As the
table above shows,
\tool{LoAT}
proves non-ter\-mi\-na\-tion  more of\-ten than
any other tool.
Accord\-ing
to the second last row, it solves 22 examples where all other
tools fail.
  Together, \tool{T2} and all \tool{TermComp} participants succeed on 1130 examples.
So \tool{LoAT} solves 23.9\% of the 92 remaining \emph{potentially} non-terminating examples.

The \tool{TPDB} examples mostly use linear arithmetic and \tool{T2} and
\tool{VeryMax} are restricted to such programs \cite{t2-tool,larraz14}.
To evaluate \tool{LoAT}
on examples with non-linear arithmetic, we also compared with the tool \tool{Anant} \cite{anant}, which has been
specifically designed to handle non-linearity. Here, we used the 29 non-terminating programs with non-linear arithmetic from the
evaluation of \cite{anant}. As we were not able to run \tool{Anant}, even though
the authors kindly provided the source code and old binaries, we compared with the results presented in
\cite{anant}.

\noindent
\begin{tabularx}{0.49\textwidth}{|c||Y|Y|Y|}
  \hline
              & \tool{LoAT} & \tool{Anant} \\ \hline \hline
  NO          &          24 &           25 \\ \hline
  MAYBE       &           5 &            4 \\ \hline
  Unique NO      &           4 &            5 \\ \hline
  Avg.\ time  &        0.5s &        32.5s \\ \hline
\end{tabularx}

Together,
\tool{Anant} and \tool{LoAT} prove non-termination of all examples.  \tool{LoAT} solves one example less than
\tool{Anant}, but it is significantly faster: It always terminates within less than three seconds whereas
\tool{Anant} takes up to 4 minutes in some cases.
However, both tools were run on different machines.

Finally,
we compared \tool{LoAT} with the
tools
from the category \emph{Termination of  \textsf{C} Integer Programs} at
\tool{TermComp}~'19\footnote{\tool{Ultimate} and \tool{AProVE} were also the two
  most powerful tools in the ``termination'' category for \textsf{C} programs at
  \tool{SV-COMP}~'19 \cite{SV-COMP19}.}
(\tool{AProVE} \cite{tool-jar}, \tool{Ultimate} \cite{Ultimate}, and \tool{VeryMax}
\cite{borralleras17,larraz14})
on the 355 examples from that category of the \textsf{TPDB}.
As \tool{LoAT} cannot parse \textsf{C}, we coupled it with a version of  \tool{AProVE} that converts \textsf{C} programs into equivalent
integer programs.

\noindent
\begin{tabularx}{0.49\textwidth}{|c||Y|Y|Y|Y|Y|}
  \hline
              & \tool{LoAT} & \tool{AProVE} &      \tool{Ultimate} & \tool{VeryMax} \\ \hline \hline
  NO          &          96 &            99 &              88 &            102 \\ \hline
  YES         &           0 &           214 &             206 &            212 \\ \hline
  MAYBE       &         239 &            22 &              41 &             21 \\ \hline
  Unique NO      &           2 &             0 &               0 &              2 \\ \hline
  Avg.\ time  &        3.1s &          6.3s &            8.7s &           5.2s \\ \hline
\end{tabularx}

 The results of \tool{LoAT} are competitive, but it succeeds on less examples than \tool{AProVE} and
\tool{VeryMax}. \tool{VeryMax} and \tool{LoAT} are the only tools that find unique non-termination proofs.
  Finally, \tool{LoAT} is the fastest
tool, although its runtime includes \tool{AProVE}'s conversion from \textsf{C}. However,
all tools but \tool{LoAT} also spend time on attempting to prove termination,
which may explain their longer runtime.

To explain the discrepancy between the results for integer programs and for
\textsf{C} programs, note that the integer programs from the \textsf{TPDB} often contain
several loops. Here,
our loop acceleration technique is particularly successful, because the challenge is not only to
prove non-termination of one of the loops, but also to prove its reachability.
In contrast,
many
\textsf{C} programs from the \textsf{TPDB} consist of a single multi-path loop. So to
prove non-termination, one has to find a suitable pattern to execute the paths through the loop's body.
To improve the handling of such examples, we will extend our approach by \emph{control flow refinement} techniques
\cite{speed-pldi-09,cfrfWST,cofloco2,underapprox15} in future work.

See \url{https://ffrohn.github.io/acceleration} for a pre-compiled binary (Linux, 64 bit) of \tool{LoAT},
tables with detailed results for all benchmarks, and the full output
of the tools
  for all
examples (the detailed results of \tool{Anant} can be found in \cite{anant}). The source code
of the implementation in our tool \tool{LoAT} is available at
\url{https://github.com/aprove-developers/LoAT/tree/nonterm}.

\section{Conclusion and Related Work}
\label{sec:conclusion}

\subsection{Conclusion}

We presented the first
non-termination technique based on loop acceleration. It accelerates terminating
loops in order to prove reachability of non-terminating configura\-tions,  even if this
requires reasoning
about program parts that contain loops themselves. As we use a non-termination criterion which is a special case of the
prerequisites of our novel loop acceleration technique (see \Cref{sec:simplify}), we can use the same new invariant
inference technique (\Cref{sec:inv}) to facilitate both loop acceleration
and non-termina\-tion proving. The experimental evaluation of our approach shows that it is competitive with state-of-the-art tools,
cf.\ \Cref{sec:experiments}.

\smallskip

\subsection{Related Work}

Loop Acceleration is mostly used in over-approximating settings (e.g., \cite{kincaid15, gonnord06, jeannet14,
  madhukar15}), whereas our setting is under-approximating. We only know of two other
under-approximating loop acceleration techniques \cite{ijcar16,underapprox15}:
One requires \emph{metering functions} \cite{ijcar16}, an adaption of ranking functions,
that can be challenging to
synthesize. The other \cite{underapprox15} is a special case
of \Cref{thm:loop-acceleration} where $\phi_{ci} = \phi_{si} = \true$, which
restricts its applicability in comparison to our approach.
To facilitate acceleration, \cite{underapprox15} splits disjunctive guards, which is
orthogonal to our splitting of guards by adding conjuncts
(cf.\ \Cref{sec:local-invariants}).

Most techniques to prove non-termination first generate \emph{las\-sos}
consisting of a simple loop $\alpha$ and a \emph{stem}, i.e., a
path from the program's entry point to
$\alpha$. Then they try to prove non-termination of these lassos.
However, a program with consecutive or nested loops usually
has infinitely many possible lassos. In contrast, our program simplification
framework yields a loop-free simplified program with finitely many transitions.

In  \cite{rupak08}, \emph{recurrent sets} were proposed to prove
non-termina\-tion. A set of configurations
is recurrent if each element has a successor in the set.
  Hence, a
non-empty
recurrent set that contains an initial configuration witnesses non-termination.
There are many techniques to find recurrent sets for simple loops \cite{amram18}, lassos \cite{rupak08,jbc-nonterm,t2-nonterm,velroyen},
or more complex sub-programs \cite{larraz14}.
Essentially, our invariant inference
technique of \Cref{sec:inv} also searches for a recurrent set for a simple loop $\alpha$.
However, if it cannot find
a recurrent set it may still successfully enforce monotonicity of $\guard{\alpha}$ and hence allow us to accelerate
$\alpha$.

An alternative to recurrent sets is presented in \cite{geometricNonterm}. It represents infinite runs
as sums of geometric series.
In general, we could use any technique to prove non-termination of simple loops or lassos
as an alternative to our non-termination criteria.

Further approaches to prove non-termination are, e.g., based on Hoare-style reasoning \cite{hiptnt} or
safety proving \cite{seahorn-term}.

While most related techniques to prove non-termination focus on
linear arithmetic, \cite{anant} has been specifically designed to handle non-linear arithmetic
via \emph{live abstractions}
and a variation of recurrent sets.
As shown in \Cref{sec:experiments},
our approach is also competitive on programs with non-linear arithmetic.

\smallskip

\subsection{Future Work}
\label{sec:future}

We will integrate control flow refinement techniques and more powerful non-termination
criteria (e.g., to find disjunctive recurrent sets, which we cannot handle yet).
We will also consider techniques to infer non-linear invariants, as our current invariant inference is restricted to linear arithmetic.

\smallskip

\subsection{Acknowledgments} We thank Marc Brockschmidt and Matthias Naaf for important
initial discussions.

\printbibliography

\newpage

\appendix

\end{document}